\documentclass[conference]{IEEEtran}
\usepackage[backend=bibtex,style=ieee]{biblatex}
\usepackage{amsmath,amssymb,amsfonts}

\usepackage{soul}
\soulregister\cite7
\usepackage{cleveref}
\usepackage{amsmath,amsfonts,amssymb}
\usepackage{booktabs}
\usepackage{algorithm}
\usepackage{algorithmic}
\usepackage{array}
\usepackage[caption=false,font=normalsize,labelfont=sf,textfont=sf]{subfig}
\usepackage{textcomp}
\usepackage{stfloats}
\usepackage{url}
\usepackage{verbatim}
\usepackage{graphicx}
\usepackage[style=ieee,backend=bibtex]{biblatex}
\usepackage{wrapfig}
\usepackage{tabularx}
\usepackage{enumitem,kantlipsum}
\DeclareFieldFormat{doi}{} 
\usepackage{cleveref}
\crefname{figure}{Fig.}{Figs.}
\Crefname{figure}{Fig.}{Figs.}
\crefname{equation}{Eq.}{Eqs.}
\Crefname{equation}{Eq.}{Eqs.}
\usepackage{xcolor}
\crefname{figure}{Fig.}{Figs.}
\Crefname{figure}{Fig.}{Figs.}
\crefname{equation}{Eq.}{Eqs.}
\Crefname{equation}{Eq.}{Eqs.}
\usepackage{algorithmic}
\usepackage{algorithm}
\usepackage{caption}
\usepackage{textcomp}
\usepackage{xcolor}
\def\BibTeX{{\rm B\kern-.05em{\sc i\kern-.025em b}\kern-.08em
    T\kern-.1667em\lower.7ex\hbox{E}\kern-.125emX}}
\DeclareFieldFormat{doi}{} 

\begin{document}

\title{
Sparsity-Aware Near-Field Beam Training via Multi-Beam Combination
}
\author{
\IEEEauthorblockN{Zijun Wang\IEEEauthorrefmark{1}, Rama Kiran\IEEEauthorrefmark{2}, Jinesh Nair\IEEEauthorrefmark{2}, Chien-Hua Chen\IEEEauthorrefmark{3},Tzu-Han Chou\IEEEauthorrefmark{4}, Shawn Tsai\IEEEauthorrefmark{4}, Rui Zhang\IEEEauthorrefmark{1}}
\IEEEauthorblockA{\IEEEauthorrefmark{1}Department of Electrical Engineering, The State University of New York at Buffalo, Buffalo, NY, USA\\ Emails: {zwang267@buffalo.edu, rzhang45@buffalo.edu}}
\IEEEauthorblockA{\IEEEauthorrefmark{2}CSD, MediaTek Inc., Bengaluru, India.  Emails: rama.kiran@mediatek.com, Jinesh.Nair@mediatek.com.}
\IEEEauthorblockA{\IEEEauthorrefmark{3}CSD, MediaTek Inc., Hsinchu, Taiwan. Email: CHBen.Chen@mediatek.com.}
\IEEEauthorblockA{\IEEEauthorrefmark{4}CSD, MediaTek Inc., San Diego CA, USA. Emails: shawn.tsai@mediatek.com, tzu-han.chou@mediatek.com}
}

\maketitle

\begin{abstract}
This paper proposes an adaptive near-field beam training method to enhance performance in multi-user and multipath environments. The approach identifies multiple strongest beams through beam sweeping and linearly combines their received signals—capturing both amplitude and phase—for improved channel estimation. Two codebooks are considered: the conventional DFT codebook and a near-field codebook that samples both angular and distance domains. As the near-field basis functions are generally non-orthogonal and often over-complete, we exploit sparsity in the solution using LASSO-based linear regression, which can also suppress noise. Simulation results show that the near-field codebook reduces feedback overhead by up to 95\% compared to the DFT codebook. The proposed LASSO regression method also maintains robustness under varying noise levels, particularly in low SNR regions. Furthermore, an off-grid refinement scheme is introduced to enhance accuracy especially when the codebook sampling is coarse, improving reconstruction accuracy by 69.4\%.
\end{abstract}

\begin{IEEEkeywords}
 Beam training, near-field communication, Terahertz communication, beamforming
\end{IEEEkeywords}

\section{Introduction}
\label{sec:intro}
\IEEEPARstart{T}{he} next generation of wireless systems is pushing ever higher in frequency—from sub-6 GHz (FR1) through upper-midband (FR3, 7-24 GHz), mmWave (FR2, 24-52 GHz) and mmWave II (FR4, 71-114 GHz) up to the THz band (0.1–10 THz), enabling immersive applications such as augmented and virtual reality (AR/VR)\cite{10438977,9766110}. To capitalize on these high-frequency opportunities, next-generation systems are expected to deploy extremely large antenna arrays (ELAAs) that promise enhanced coverage and significantly boosted data throughput. However, the convergence of large antenna apertures and shorter wavelengths gives rise to near-field effects that challenge conventional far-field assumptions \cite{General_tutorial2, General_tutorial3}. In the near-field regime, electromagnetic wavefronts exhibit spherical rather than planar characteristics, making the steering vector sensitive to both the angle and distance of the target user \cite{The_Los_dominated_Tutorial, Fraunhofer_and_Fresnel_Distances}. This fundamental shift calls for innovative beamforming and beam training strategies tailored specifically for near-field users (NUs).

Recent research has addressed this challenge by proposing specialized codebooks and hierarchical training methods. For example, in \cite{Cui_and_Dai_channel_model}, a near-field codebook in the polar domain that integrates both angular and distance information is developed for uplink channel estimation in Time Division Duplex (TDD) systems but is unsuitable for Frequency Division Duplex (FDD) operation. 
 Building on the near-field codeook proposed in \cite{Cui_and_Dai_channel_model}, two-stage near-field beam training schemes have been introduced in FDD systems to reduce the complexity of exhaustive searches using the near-field codebook \cite{Two-Stage_Hierarchical_Beam_Training, Fast_Near-Field_Beam_Training, Efficient_Hybrid_Near-_and_Far-Field_Beam_Training}. These methods utilize DFT codewords for beam sweeping to initially pinpoint the angular-domain location, and then refine the estimation within the angular or both angular and distance domains, based on user feedback of the beam index that delivers the highest received power. Additionally, \cite{Joint_Angle_and_Range_Estimation} analyzes the received signal beam pattern to locate users and selects the corresponding polar-domain codeword for beamforming. Although these schemes apply to FDD systems, they only select the strongest or line-of-sight (LoS) path for beamforming through beam sweeping, which can lead to energy loss and performance degradation in multipath scenarios—especially in multi-user systems.

In the far-field scenario, 5G new radio (NR) introduced Type II codebook framework.  The framework adopts a linear combination of the selected DFT codewords with the strongest received power from user feedback, thereby utilizing significant multipath components for data transmission. The Type II codebook demonstrates advantages such as multibeam support and precise channel feedback, making it particularly effective in multipath environments and multi-user MIMO (MU-MIMO) systems \cite{10304476,qin2023reviewcodebookscsifeedback}. Nevertheless, the Type II codebook is fundamentally designed under the far-field channel model assumption. Recent efforts to extend its applicability to near-field scenarios have led to an enhanced scheme, which employs a weighted superposition of multiple DFT vector-based beams \cite{10768093}. However, this approach only considers LoS path and fails to address its applicability in multipath environments. Additionally, the DFT codebook suffers from high feedback overhead in the near-field region due to the "energy-split" effect, where beam energy disperses across multiple spatial dimensions \cite{Cui_and_Dai_channel_model,Joint_Angle_and_Range_Estimation}.

In this paper, we propose a multi-beam combining framework to address multipath propagation in near-field environments. The base station (BS) performs beam sweeping using various codewords, and the users feedback the $K$ beam indices that yield the strongest received powers, along with the corresponding amplitude and phase information. The BS then estimates the channel by weighting and combining these selected beams based on the feedback.  We analyze and benchmark two codebooks, the conventional DFT codebook and the near-field codebook. We further introduce a sparse reconstruction method using the least absolute shrinkage and selection operator (LASSO), which not only identifies dominant propagation paths but also suppresses noise. Furthermore, we enhance estimation accuracy through an off-grid refinement process based on LASSO, mitigating discretization errors caused by the finite resolution of the codebook, especially when the codebook size is small.

The rest of the paper is organized as follows. \Cref{sec:system model} presents the system model and the associated signal model. \Cref{sec:Proposed Scheme} proposes the multi-beam linear combination scheme. \Cref{sec:simulation} analyzes the framework's performance through numerical simulations and benchmarks the DFT and near-field codebooks.  \Cref{sec:RefinedLASSO} presents an off-grid refinement method and \cref{sec:clun} concludes the work. 

\section{System Model}\label{sec:system model}
We consider a scenario in which a BS is furnished with a uniform linear array (ULA) containing \(N\) antenna elements, while each user is equipped with a single antenna.
In this work, we adhere to the definition provided in \cite{Fraunhofer_and_Fresnel_Distances} by defining the near-field region to be bounded by the Fresnel distance $R_{\text{Fre}} = \frac{1}{2}\sqrt{\frac{D^3}{\lambda}}$ and the Rayleigh distance $R_{\text{Ray}} = \frac{2D^2}{\lambda}$, where \(\lambda\) is the center wavelength and the ULA aperture given as \(D = (N-1)d\).
 
\subsection{Channel Model}\label{subsec:Channel Model}
\begin{figure}[!t]
\captionsetup{justification=justified,singlelinecheck=false} 
\centering
\includegraphics[width=0.8\linewidth]{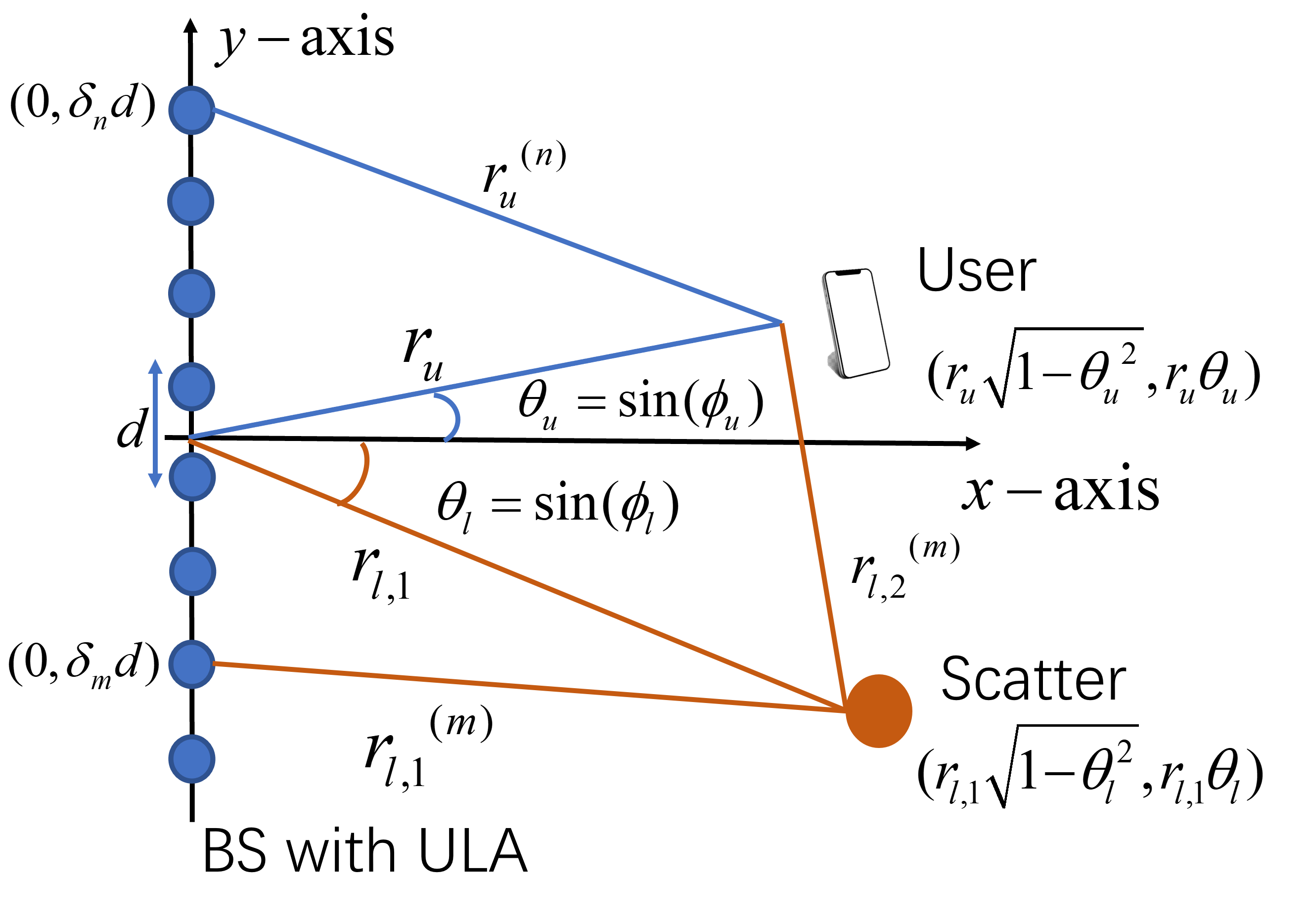}
\caption{Diagram of the near-field communication system featuring a ULA.}
\label{fig:antenna}
\end{figure}

An illustration of our channel model is presented in \Cref{fig:antenna}. We assume the antenna array is oriented along the \(y\)-axis and centered at the origin \((0,0)\). The coordinate for the \(n\)-th antenna element is defined as \((0, \delta_n d)\), where $\delta_n = \frac{2n-N+1}{2},n \in \mathcal{N}, \text{and}\;\mathcal{N} \triangleq \{0,1,\dots, N-1\}.$  
The spacing between antennas is set as \(d = \frac{\lambda}{2}\). The near-field steering vector is given by 
\begin{equation}
     \mathbf{b}(\theta, r) = \frac{1}{\sqrt{N}}\Biggl[ e^{-j\frac{2\pi (r^{(0)}-r)}{\lambda}},\ldots, e^{-j\frac{2\pi (r^{(N-1)}-r)}{\lambda}} \Biggr]^{T},
     \label{eq:near_field_steering_vector}
\end{equation}
where the spatial parameter is defined as \(\theta = \sin(\phi)\) (with \(\phi\) being the angle of departure) and lies within the interval \([-1,1]\). The distance from the \(n\)-th antenna element to the NU or scatter is expressed as $r^{(n)} = \sqrt{r^{2} + \delta_n^2 d^{2} - 2r\,\theta\, \delta_n d}$, with \(r\) denoting the distance between the center of the antenna array and the user. 

A multipath effect of total $L$ paths is considered. For LoS part, the channel between the BS and the NU located at $(\theta_u,r_u)$ is modeled as 
\begin{equation}
   \mathbf{h}_{\text{LoS}}= \sqrt{N} g_u e^{-j\frac{2\pi r_u}{\lambda}} \mathbf{b}(\theta_u, r_u),
   \label{eq:LoS channel_model}
\end{equation}
where \(j = \sqrt{-1}\) and $g = \frac{\lambda}{4\pi r_u}$ represents the channel gain due to the path loss between the BS and the user. And for non-line-of-sight (NLoS) channel $\mathbf{h}_{\text{NLoS}}$, we have
\begin{equation}
   \mathbf{h}_{\text{NLoS}}= \sqrt{N}\sum_{l=1}^{L-1} g_l e^{-j\frac{2\pi (r_{l,1}+r_{l,2})}{\lambda}} \mathbf{b}(\theta_l, r_{l,1}),
   \label{eq:NLoS channel_model}
\end{equation}
where $l$-th scatter is set at location $(\theta_l,r_{l,1})$. $r_{l,2}=\sqrt{r_{l,1}^{2} + r_u^{2} - 2r_ur_{l,1}\cos(\theta_u-\theta_l)}$ according to law of cosines is the distance from scatter to user. $g_l=\frac{\lambda p_l}{4\pi (r_{l,1}\cdot r_{l,2})}$ is the channel gain that includes the path loss and the random reflection character of $l$-th scatter $p_l\sim\mathcal{CN}(0,1)$. Then, the channel model consists of multiple paths is written as
\begin{equation}
       \mathbf{h}=\mathbf{h}_{\text{LoS}}+\mathbf{h}_{\text{NLoS}}
    \label{eq:channel model}
\end{equation}
\subsection{Received Signal}\label{subsec:signal}
Based on the channel model, the received signal $y(\mathbf{v})$ at a user located at $(\theta, r)$ is expressed as:
\begin{equation}
  y(\mathbf{v})=\mathbf{h}^{H}\mathbf{v} x+w,  
  \label{eq:received signal model}
\end{equation}
where $\mathbf{v}$ represents the beamforming vector, and $w\sim \mathcal{CN}(0,\sigma^2)$ is the complex additive white Gaussian noise (AWGN). The reference pilot signal is denoted as $x=1$ without loss of generality.

For the DFT codebook $\mathbf{V}_{DFT}=\{\mathbf{a}(\varphi_0),\mathbf{a}(\varphi_1),\cdots,\mathbf{a}(\varphi_{N-1})\}$, the angles are uniformly sampled, i.e. $\varphi_n = \frac{2n - N + 1}{N}$ for $n \in \mathcal{N}$ and corresponding beamforming vector in the far-field is given by:
\begin{align}
 &\mathbf{a}(\varphi_n) \nonumber\\
 &\triangleq \frac{1}{\sqrt{N}} 
\left[ 
e^{-j\pi \left(-\frac{N-1}{2}\right)\varphi_n},
\cdots,e^{-j\pi \left(\frac{N-1}{2}\right)\varphi_n}
\right]^T,\forall n \in \mathcal{N}.   
 \label{eq:DFT codeword}
\end{align}
For the near-field codebook, we adopt the approach presented in \cite{Cui_and_Dai_channel_model}. This method constructs the codebook in the polar domain by uniformly sampling the angle and non-uniformly sampling the distance. The sampling strategy is designed to minimize the correlation between adjacent codewords—whether in terms of angle or distance. For a codeword corresponding to the position $(\varphi_s,r_s)$, the Taylor expansion $r_s^{(n)}\approx r_s-\delta_{n} d \varphi_s+\frac{\delta_{n}^{2} d^{2}\left(1-\varphi_s^{2}\right)}{2 r_s}$ is applied, and thus the codeword can be simplified from \Cref{eq:near_field_steering_vector} as

{\small
\begin{align}
\mathbf{ b}'(\varphi_s,r_s)
&= \frac{1}{\sqrt{N}}\Bigl[
\exp\Bigl(-\jmath\Bigl(\pi\,\delta_{0}\,\varphi_s
      +\frac{\pi\,d\,\delta_{0}^{2}(1-\varphi_s^{2})}{2\,r_s}\Bigr)\Bigr),\dots,\nonumber\\
&\qquad\quad
\exp\Bigl(-\jmath\Bigl(\pi\,\delta_{N-1}\,\varphi_s
      +\frac{\pi\,d\,\delta_{N-1}^{2}(1-\varphi_s^{2})}{2\,r_s}\Bigr)\Bigr)
\Bigr]^T.
\label{eq:near-field codeword}
\end{align}
}

\section{Linear-combination Scheme}\label{sec:Proposed Scheme}
\begin{figure}[!t]
\captionsetup{justification=raggedright, singlelinecheck=false} 
\centering
\subfloat[\tiny(a)][\textrm{\small Far-field codebook with 512 codewords}]{%
    \includegraphics[width=0.8\linewidth]{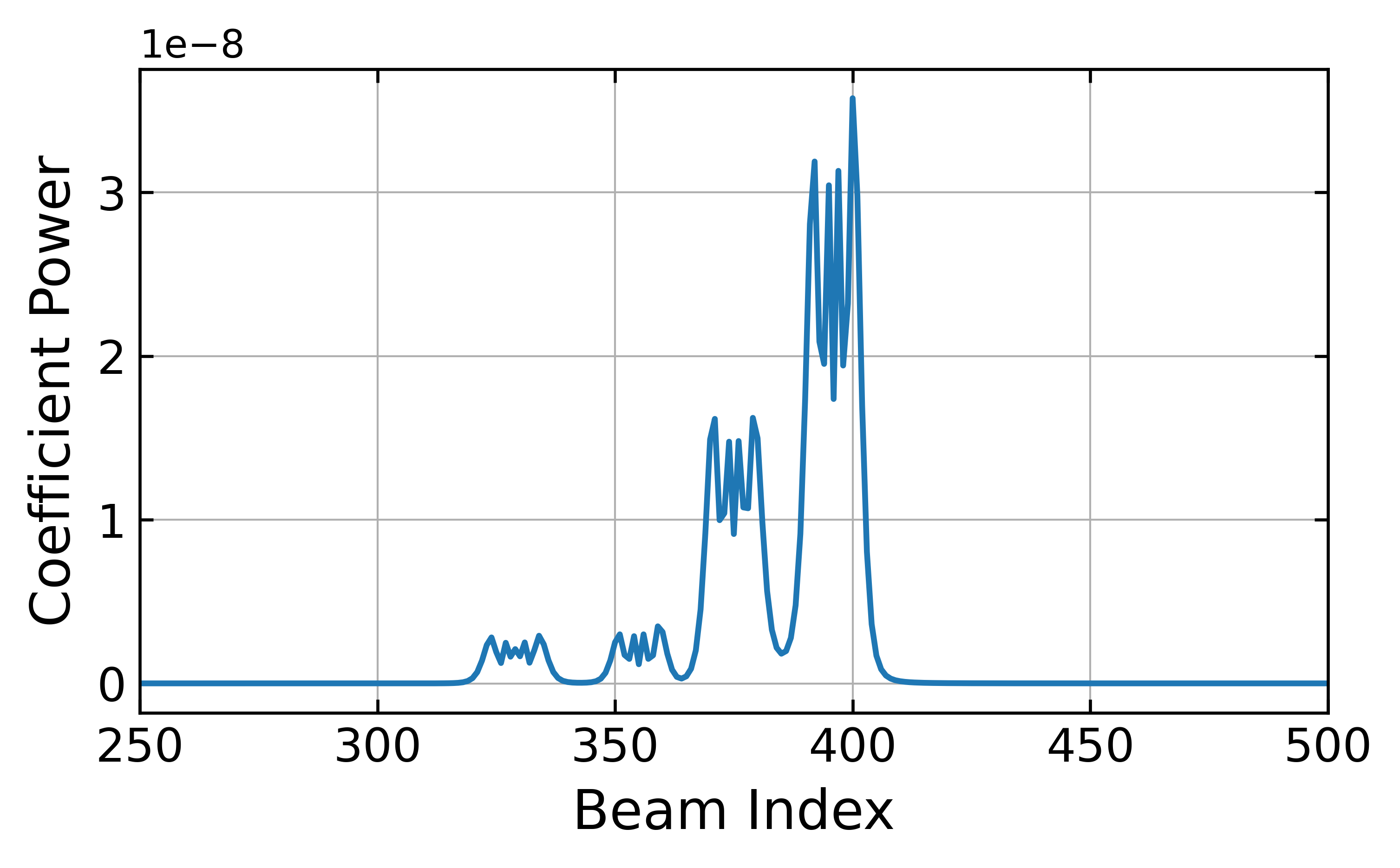}%
    \label{fig:far_index}%
}\\
\subfloat[\tiny(a)][\textrm{\small Near-field codebook with 1890 codewords.}]{%
    \includegraphics[width=0.8\linewidth]{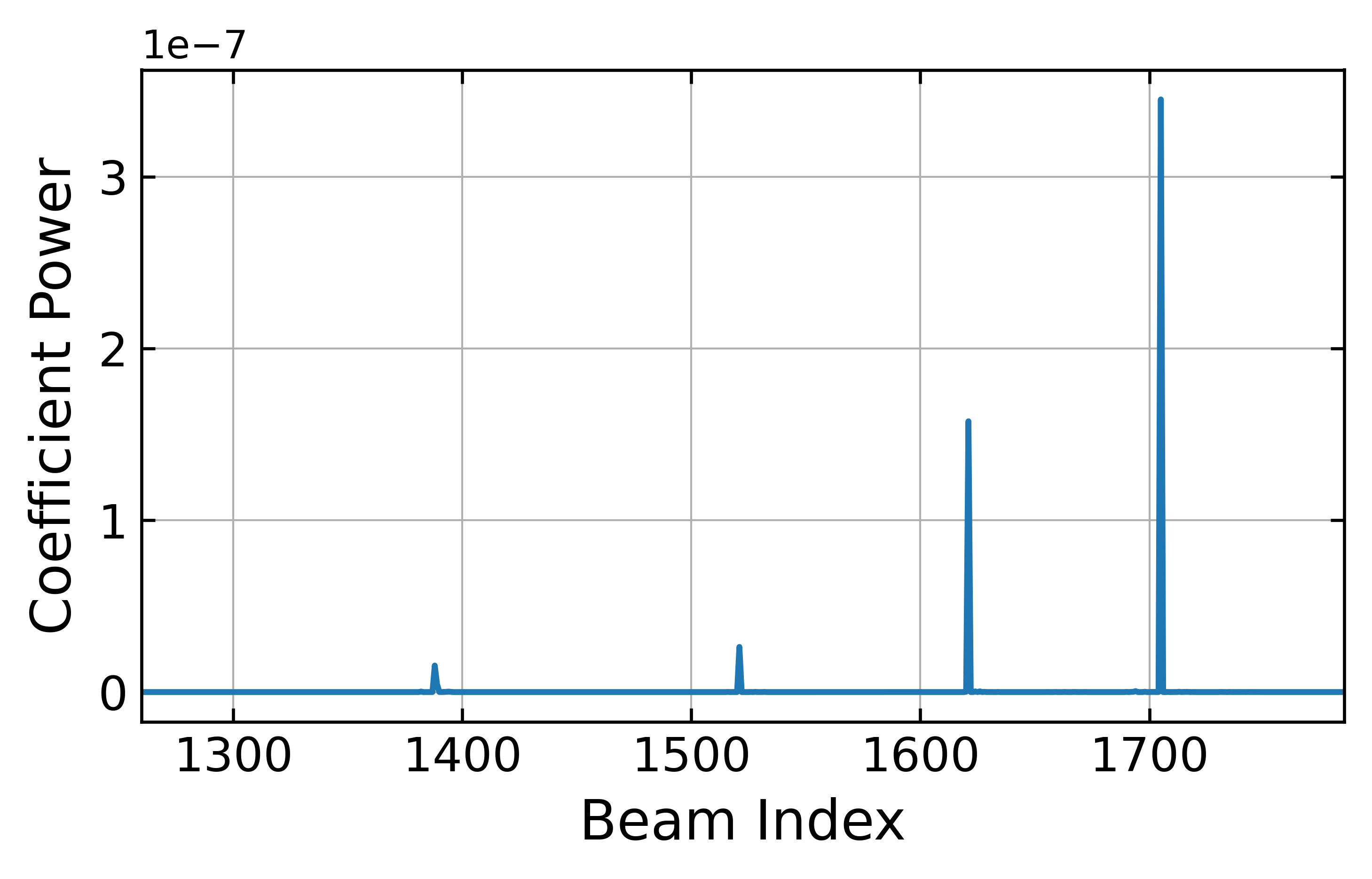}%
    \label{fig:near_index}%
}
\caption{Coefficient power versus codebook beam index.}
\label{fig:beam index}
\end{figure}

In this section, we introduce the principle of the linear combination scheme for near-field beam training. 
The BS perform beam sweeping with a set of beam sweeping precoding vectors given by
\begin{equation}
     \mathbf{V} = \begin{bmatrix}\mathbf{v}_0, \mathbf{v}_1, \dots, \mathbf{v}_{M-1}\end{bmatrix},
     \label{eq:codebook_matrix}
\end{equation}
where $M\ge N$ denotes the total number of beams employed for beam sweeping. Assuming the pilot signal is $1$, the received signal vector at user side is expressed as  
 \begin{equation}
    \mathbf{y} = \mathbf{h}^H\mathbf{V}+\mathbf{W},
    \label{eq:receive_signal_vector}
\end{equation}
where $\mathbf{y} = \begin{bmatrix} y(\mathbf{v}_0) , y(\mathbf{v}_1) , \cdots , y(\mathbf{v}_{M-1}) \end{bmatrix}$, $y(\mathbf{v}_i)$ represents the signal measured at the user side when employing the $i$-th beam corresponding to the codeword $\mathbf{v}_i$, $i=0,1,\dots, M-1$. Given that the multipath effect can be expressed as a linear combination of the codewords, i.e.,
\begin{equation}
     \mathbf{h} = \mathbf{V}\boldsymbol{\alpha} ,
     \label{eq:linear_combination}
\end{equation}
where \(\boldsymbol{\alpha}=[\alpha_0, \alpha_1,\dots,\alpha_{M-1}]^T\) is the coefficient vector. \Cref{fig:beam index} illustrates examples of combining coefficients power plotted against beam index for a multipath near-field channel using different codebooks. As both schemes exhibit a certain level of sparsity, users can reduce feedback overhead by selecting only the $K\leq M$ beams with the highest received signal powers. The corresponding received signal amplitudes and phases for these $K$ beams are then fed back to reconstruct the near-field channel. Let $\mathcal{K}$ denote the set of selected beam indices with cardinality $K$. The corresponding sub-codebook is $\mathbf{V}_{\mathcal{K}}\in\mathbb{C}^{N\times K}$ and coefficient sub-vector is  \(\boldsymbol{\alpha}_{\mathcal{K}}\in\mathbb{C}^{K\times1}\), where \(\mathbb{C}\) denotes the field of complex numbers. In this way, \Cref{eq:linear_combination} can be further written as 
\begin{equation}
     \mathbf{h} \approx \mathbf{V}_{\mathcal{K}}\boldsymbol{\alpha}_{\mathcal{K}}.
     \label{eq:linear_combination2}
\end{equation}
Moreover, as depicted in \Cref{fig:far_index}, the DFT codebook demonstrates an energy-split effect in near-field scenarios, whereas \Cref{fig:near_index} shows that the near-field codebook exhibits a much sparser structure, potentially reducing feedback overhead (e.g., smaller $K$) compared to the DFT codebook.

Let $\mathbf{y}_{\mathcal{K}}\in\mathbb{C}^{1\times K}$ denote the received signal vector corresponding to the selected $K$ beams. Following the Least Square (LS) estimation rule, one can estimate the combination coefficients as
\begin{equation}
    \mathbf{\bar{\boldsymbol{\alpha}}_{\mathcal{K}}} \approx {(\mathbf{V}_{\mathcal{K}}^H\, \mathbf{V}_{\mathcal{K}})}^\dagger\mathbf{y}_{\mathcal{K}}^H,
    \label{eq:model_with_VVV}
\end{equation}
where $\cdot^\dagger$ denotes the pseudo-inverse. Then we can reconstruct the near field channel as:
\begin{equation}
    \bar{\mathbf{h}} \approx \mathbf{V}_{\mathcal{K}}{(\mathbf{V}_{\mathcal{K}}^H\, \mathbf{V}_{\mathcal{K}})}^\dagger\mathbf{y}_{\mathcal{K}}^H.
    \label{eq:channel_with_VVV}
\end{equation}
When $\mathbf{V}$ is an orthonormal basis (i.e., $N$ DFT basis), we have $\mathbf{V}_{\mathcal{K}}^H\, \mathbf{V}_{\mathcal{K}}=\mathbf{I}$, and one can reconstruct the channel by
\begin{equation}
    \bar{\mathbf{h}} \approx \mathbf{V}_{\mathcal{K}}\mathbf{y}_{\mathcal{K}}^H.
    \label{eq:channel_with_OVV}
\end{equation}
This results in a linear combination of multiple beams weighted by the measured received signals. However, when beam sweeping employs the near-field codebook from \cite{Cui_and_Dai_channel_model}, the matrix $\mathbf{V}_{\mathcal{K}}$ is no longer orthonormal. Consequently, an inverse matrix operation is required to reconstruct the channel, as shown in \Cref{eq:channel_with_VVV}. 

However, at low SNR where noise power is high, the inverse matrix operation in \Cref{eq:channel_with_VVV} can amplify noise, resulting in inaccurate channel reconstruction or overfitting.   To suppress the noise impact under low SNR, the base station can leverage LASSO regression methods as follows. 
\begin{equation}
\min_{\boldsymbol{\alpha}_{\mathcal{K}}} \; \frac{1}{2}\left\| \mathbf{y}^H_{\mathcal{K}} - \mathbf{V}_{\mathcal{K}}^H\mathbf{V}_{\mathcal{K}}\boldsymbol{\alpha}_{\mathcal{K}} \right\|_2^2 \;+\; \lambda \|\boldsymbol{\alpha}_{\mathcal{K}}\|_1,
    \label{eq:lasso}
\end{equation}
where \(\lambda > 0\) is a regularization parameter controlling the trade-off between data fidelity and sparsity.

The solution of non-zero coefficients \(\hat{\boldsymbol{\alpha}}_{\mathcal{K}}\) obtained from \Cref{eq:lasso} provides the desired linear-combination representation of the near-field channel. Then the near-field channel can be sparsely represented by the corresponding codebook $\mathbf{V}_{\mathcal{K}}$ with non-zero coefficients, we have
\begin{equation}
     \hat{\mathbf{h}} = \mathbf{V}_{\mathcal{K}} \hat{\boldsymbol{\alpha}}_{\mathcal{K}}.
     \label{eq:approximation_sub}
\end{equation}

\section{Performance Analysis}\label{sec:simulation}
In this section, we present benchmarks comparing different schemes. We consider the following schemes:
\begin{enumerate}
\item{ Linear combination of near-field codebook with LASSO (NF + LASSO scheme).} This scheme employs the near-field codebook proposed in \cite{Cui_and_Dai_channel_model} with $\beta=1.6$. $\beta$ is used to control the sampling density over distance, with adaptive sampling for distance applied at each angle. The codebook used for beam sweeping consists of 1890 codewords (sampling across 512 angles and 6 distances at most for each angles). Based on the amplitudes and phases of $K$ feedback beams, the base station performs linear combining via LASSO, as outlined in \Cref{eq:lasso} and \Cref{eq:approximation_sub}.
\item{Linear-combination of near-field codebook (NF scheme).} In this scheme, the same near-field codebook as in the NF + LASSO scheme is adopted for beam sweeping. The key difference lies in the combining process: the base station uses the amplitudes and phases of $K$ received signal from user feedback to perform linear combination based on \Cref{eq:channel_with_VVV}, which is given by $\bar{\mathbf{h}} \approx \mathbf{V_{\mathcal{K}}}{(\mathbf{V_{\mathcal{K}}}^H\, \mathbf{V_{\mathcal{K}}})}^\dagger\mathbf{y_{\mathcal{K}}}^H$. 
\item{Linear-combination of DFT codebook (DFT scheme).} This scheme utilizes the conventional DFT codebook for far-field users for beam sweeping. The codebook size is 512. Here, the base station performs linear combining from the user feedback to estimate channel based on \Cref{eq:channel_with_OVV} as $\bar{\mathbf{h}} \approx \mathbf{V_{\mathcal{K}}}\mathbf{y_{\mathcal{K}}}^H$.
\end{enumerate} 

These schemes are evaluated based on their ability to reconstruct the near-field channel accurately while minimizing feedback overhead and mitigating noise effects, particularly under low SNR conditions.
First we examine the influence of the size of index set $\mathcal{K}$, which determines the feedback overhead from users to BS. The number of path is set as $L=5$ (i.e. one LoS path and four NLoS paths).
The simulation sets the reference SNR as the SNR of a user at $(0, 5\text{m})$ without beamforming. The central frequency is $100\text{GHz}$ and all users are located in the near-field, whose locations are within the Rayleigh Distance.

\begin{figure}[!t]
  \centering
  \subfloat[\tiny(a)][\textrm{\small The L2 error versus overhead number.}]{%
    \includegraphics[width=0.45\linewidth]{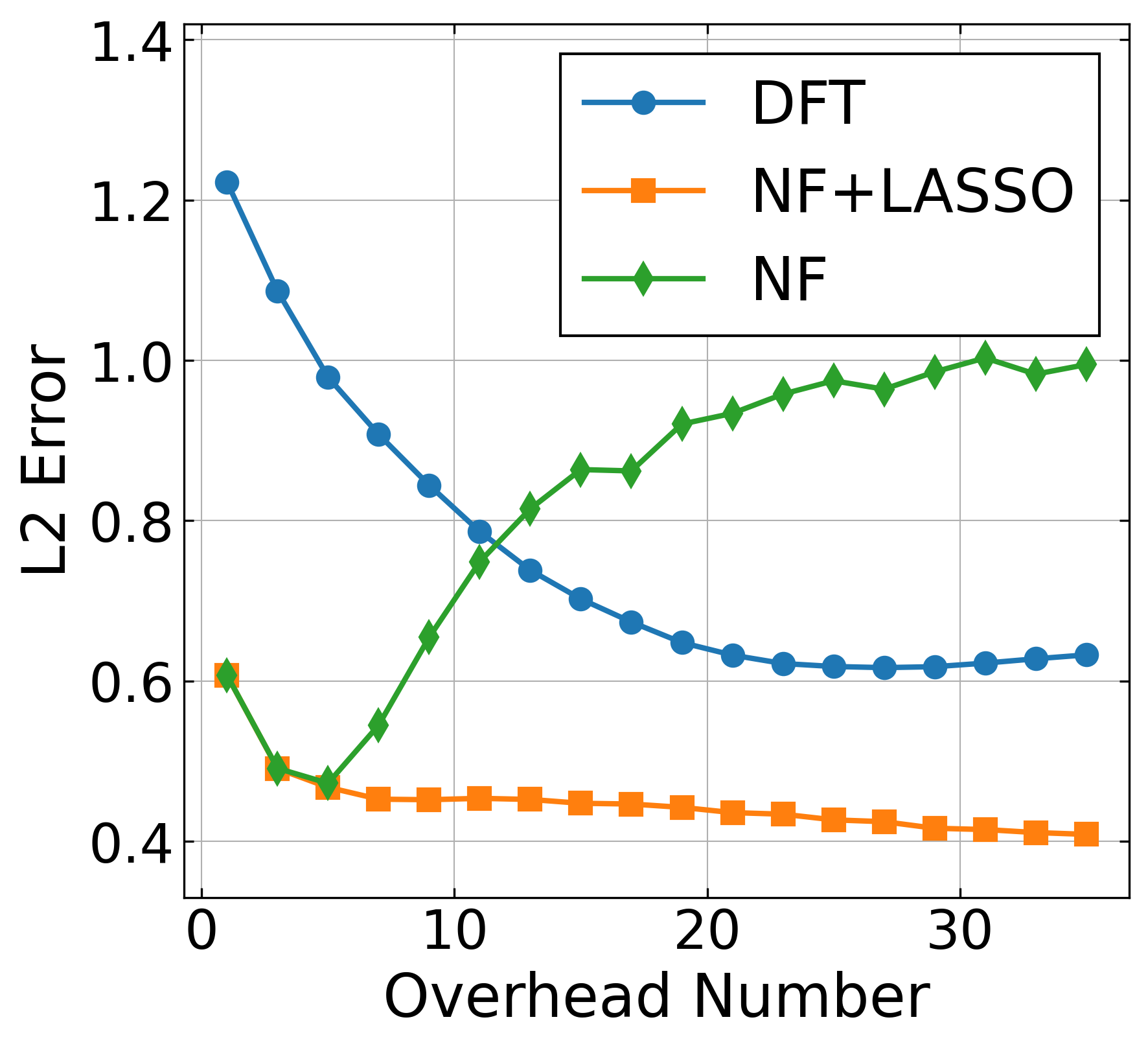}%
    \label{fig:low_overhead_l2}}%
  \quad
  \subfloat[\tiny(b)][\textrm{\small The achievable rate versus overhead number.}]{%
    \includegraphics[width=0.45\linewidth]{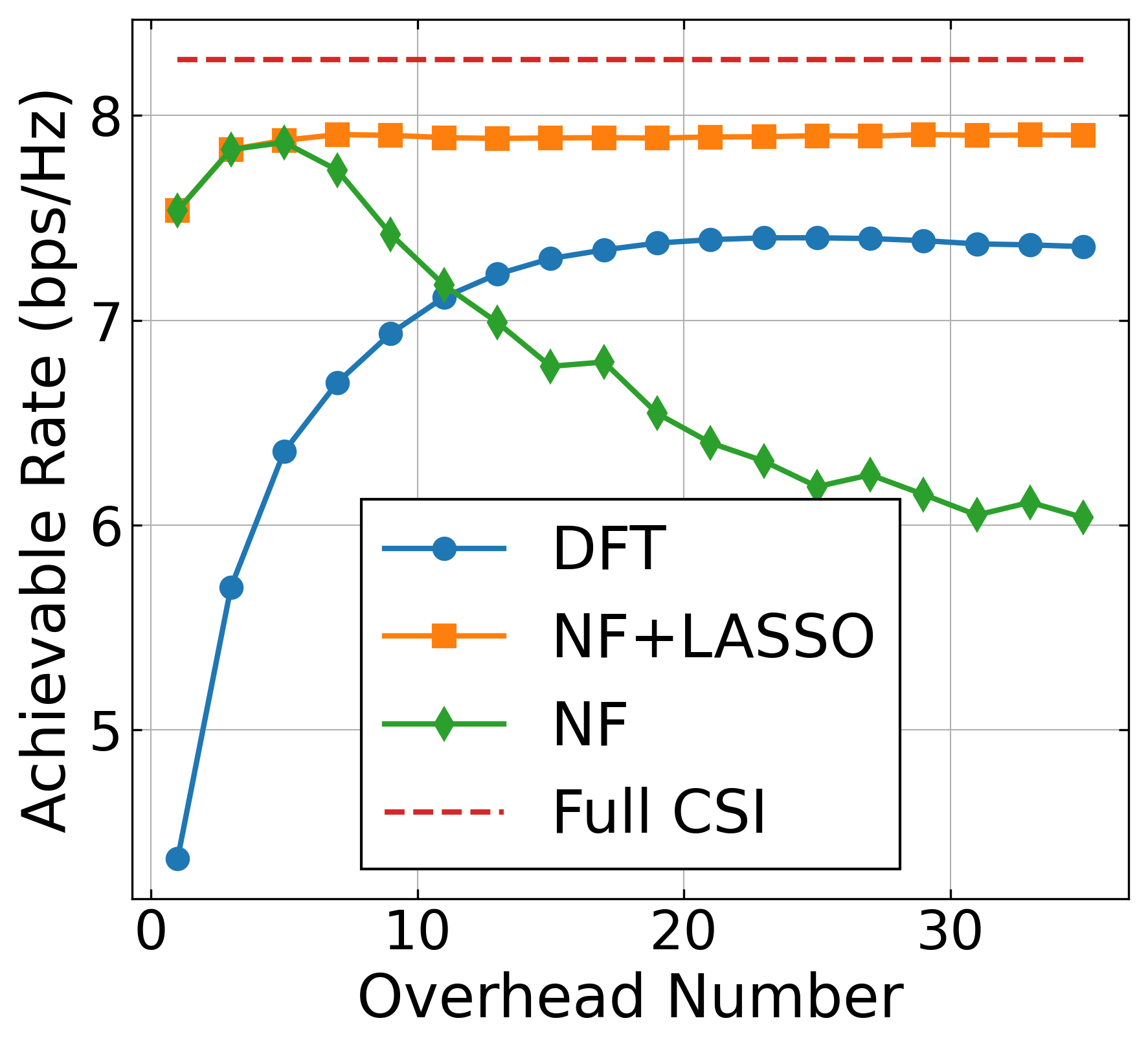}%
    \label{fig:low_overhead_rate}}%
  \captionsetup{justification=justified, singlelinecheck=false} 
  \caption{Influence of feedback overhead in a low SNR scenario.}
  \label{fig:low_snr_overhead}
\end{figure}

In the low SNR scenario (SNR = $4$ dB), we evaluate performance as illustrated in \Cref{fig:low_snr_overhead}. The results depict two key relationships: 1) The number of selected beams (i.e., the feedback overhead) versus the L2 error between the normalized estimated channel and the normalized true channel, expressed as $\|\mathbf{h}/\|{\mathbf{h}}\|_2-\hat{\mathbf{h}}/\|\hat{\mathbf{h}}\|_2\|_2$; and  2) the feedback overhead versus the achievable rate, where the achievable rate is calculated as $R=\log_2(1+\frac{\|\mathbf{h}^H\hat{\mathbf{h}}\|_2^2}{\|\hat{\mathbf{h}}\|^2_2\sigma^2})$. A red dashed line in \Cref{fig:low_overhead_rate} represents a performance upper bound, corresponding to the scenario where the true channel state information (CSI) is known at the base station. The L2 error for both the DFT and NF + LASSO schemes decreases as feedback overhead increases, due to the enhanced representation capability afforded by employing a larger number of basis elements. Consequently, the achievable rate for these schemes also increases with higher feedback overhead, as the improved accuracy of the estimated channel leads to better performance. Conversely, for the NF scheme, the L2 error initially decreases and then increases with additional feedback overhead. This behavior arises because the inverse matrix operation inherent in the NF scheme amplifies noise, as discussed in \Cref{sec:Proposed Scheme}. In contrast, the NF + LASSO scheme mitigates this noise amplification effect, resulting in a more consistent improvement with increased feedback overhead. Overall, the DFT scheme requires significantly more feedback overhead to achieve satisfactory performance. For example, to attain an achievable rate of 7.4 bps/Hz, the NF scheme and NF + LASSO scheme need a feedback overhead of only 1, whereas the DFT scheme demands 21. Moreover, the highest achievable rate of the NF + LASSO and NF schemes surpasses that of the DFT scheme. However, it is worth noting that the NF scheme is highly sensitive to the feedback overhead and performs optimally only when the overhead remains low.

\begin{figure}[!t]
  \centering
  \subfloat[\tiny(a)][\textrm{\small The L2 error versus overhead number.}]{%
    \includegraphics[width=0.45\linewidth]{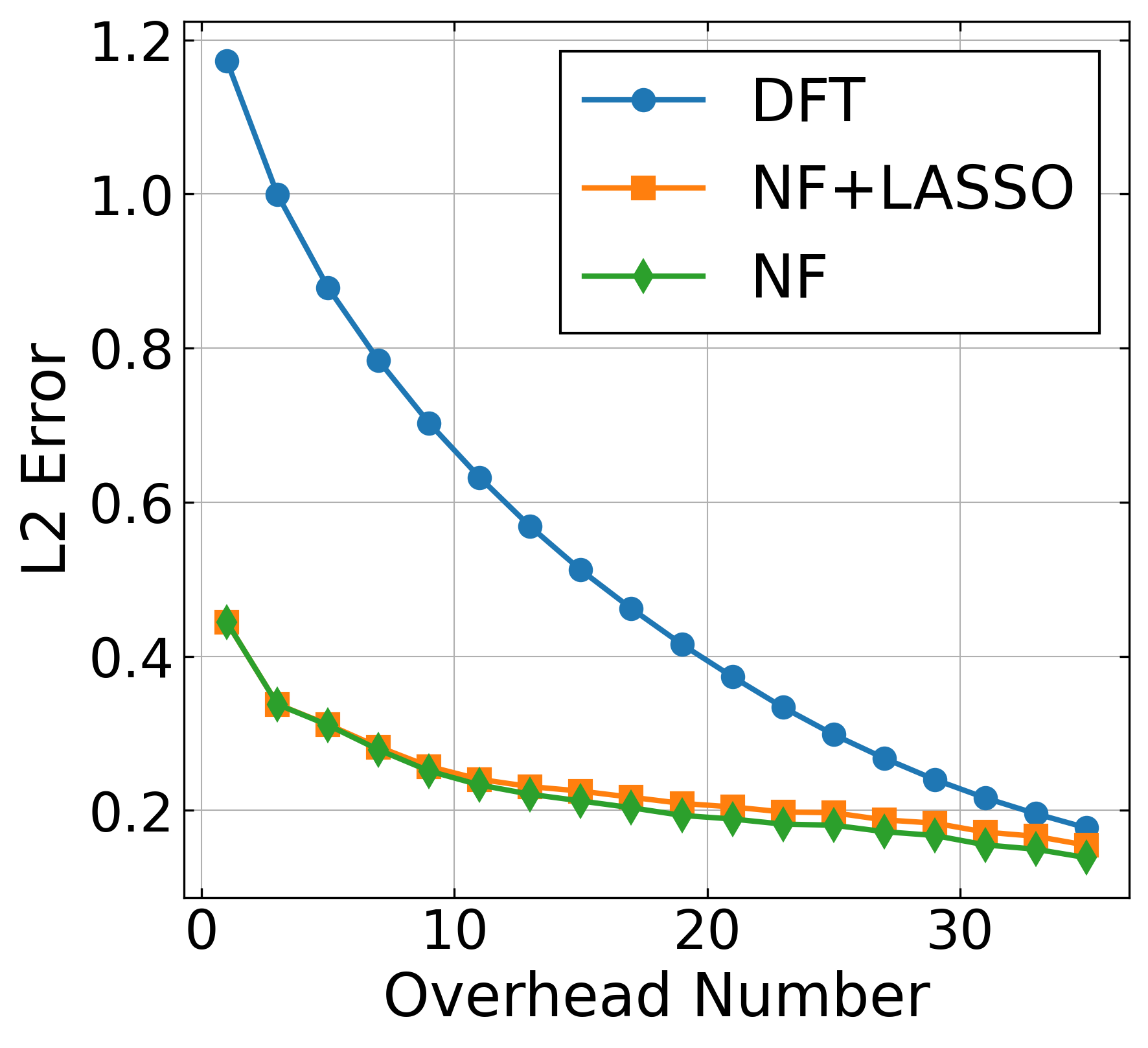}%
    \label{fig:high_overhead_l2}}%
  \quad
  \subfloat[\tiny(b)][\textrm{\small The achievable rate versus overhead number.}]{%
    \includegraphics[width=0.45\linewidth]{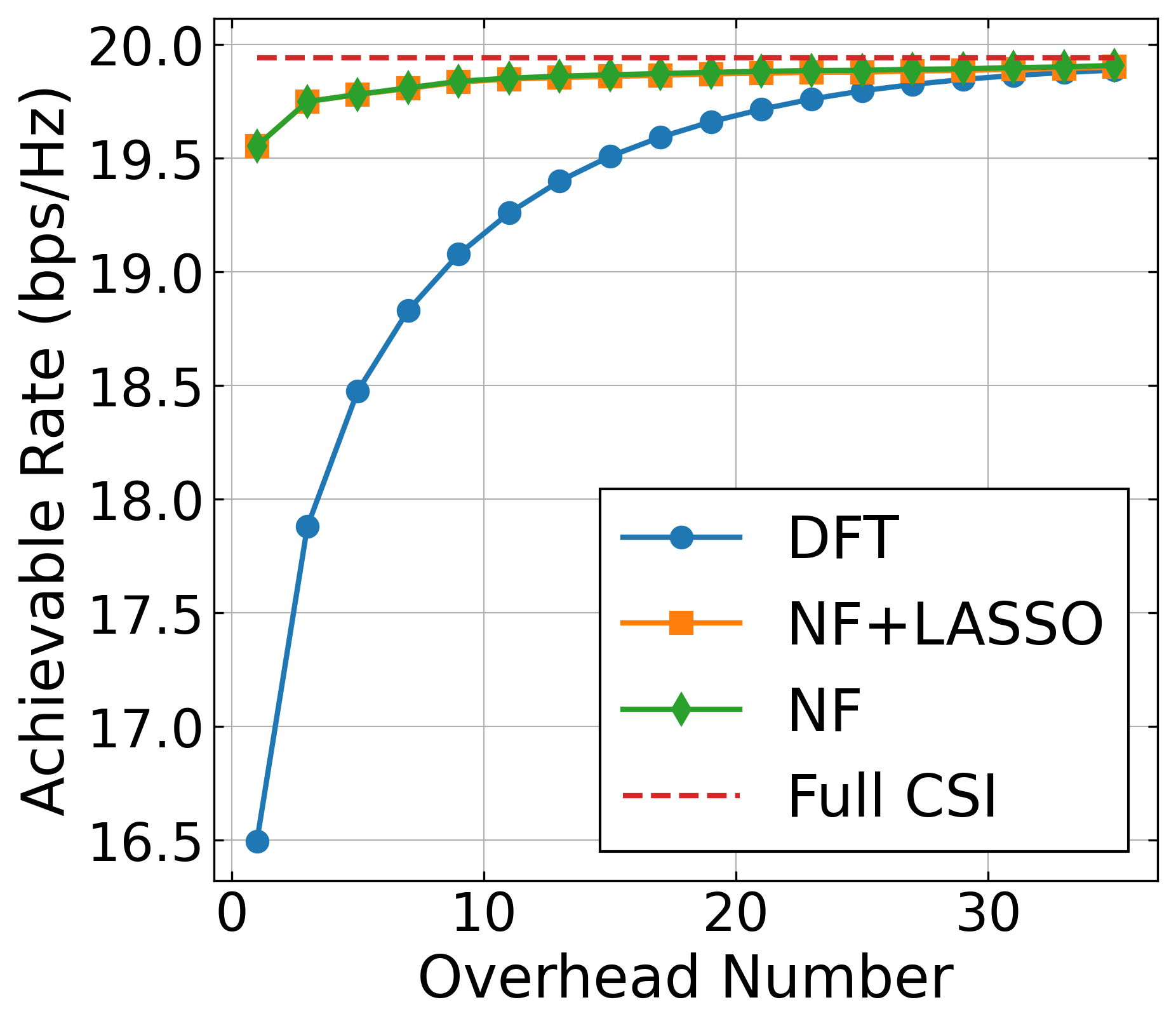}%
    \label{fig:high_overhead_rate}}%
  \captionsetup{justification=raggedright, singlelinecheck=false} 
  \caption{Influence of feedback overhead at high SNR.}
  \label{fig:high_snr_overhead}
\end{figure}

In the high SNR scenario (SNR = $40$ dB), \Cref{fig:high_snr_overhead} presents the feedback overhead versus the L2 error between the normalized estimated and true channels, as well as versus the achievable rate. Both the NF + LASSO and DFT schemes exhibit trends similar to those observed in the low SNR scenario. In contrast, the NF scheme shows improved performance at high SNR, with channel estimation accuracy and achievable rate both improving as the feedback overhead increases because the impact of noise is minimal. It is worth noting that the NF scheme performs similarly to the NF + LASSO scheme in this case. This verifies that the degradation of the NF scheme in low-SNR settings is attributable to noise amplification effects.  Nevertheless, the DFT scheme still requires considerably more feedback overhead compared to the other two schemes. For instance, to achieve an achievable rate of 19.7 bps/Hz, the DFT scheme requires a feedback overhead of 23, whereas the NF scheme attains the same rate with only 2.

\begin{figure}[!t]
  \centering
  \includegraphics[width=0.7\linewidth]{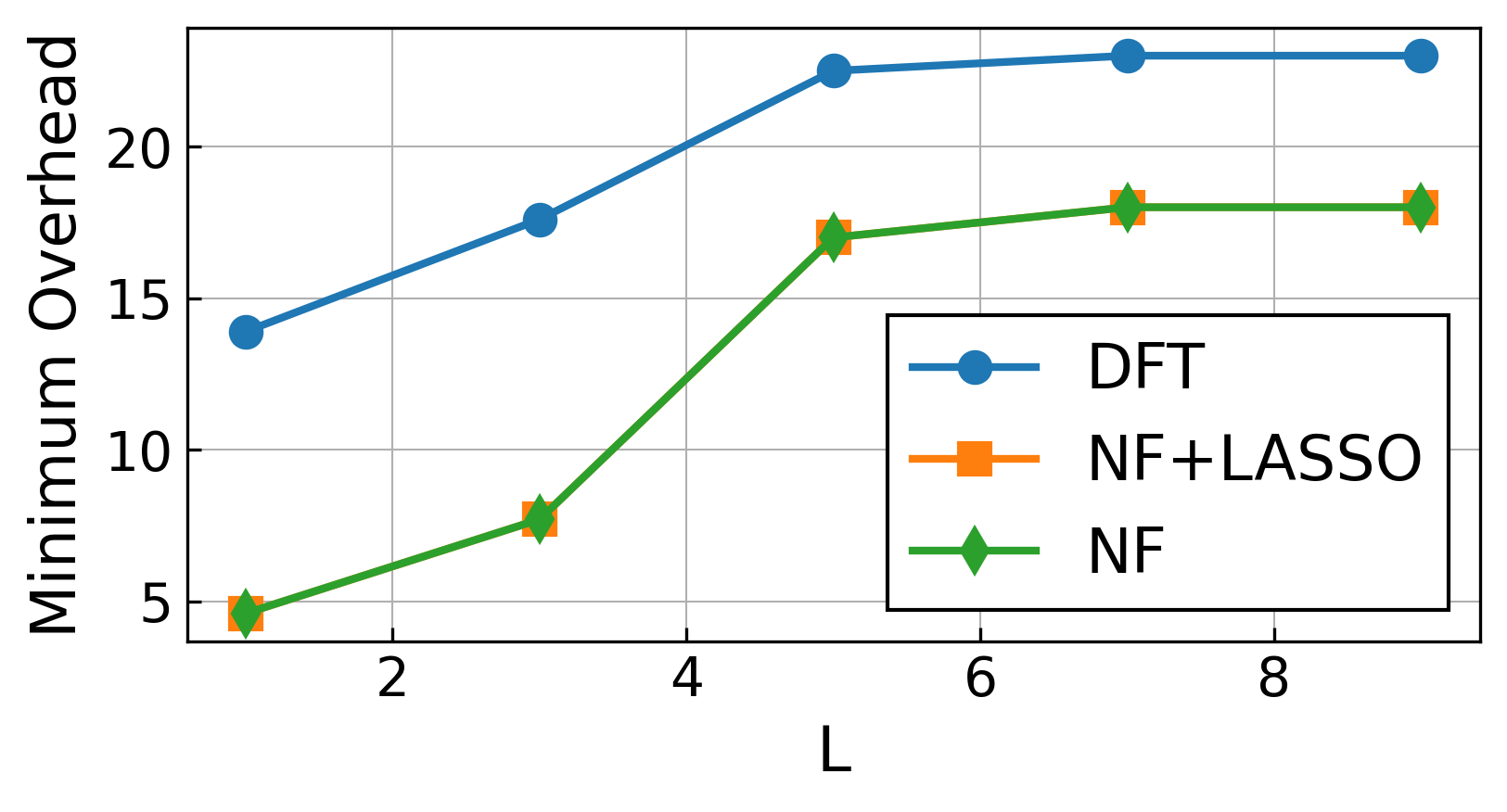}
  \captionsetup{justification=raggedright, singlelinecheck=false} 
  \caption{Influence of multipath number.}
  \label{fig:path_overhead}
\end{figure}

\begin{figure}[!t]
  \centering
  \subfloat[\tiny(a)][\textrm{\small The L2 error versus SNR.}]{%
    \includegraphics[width=0.45\linewidth]{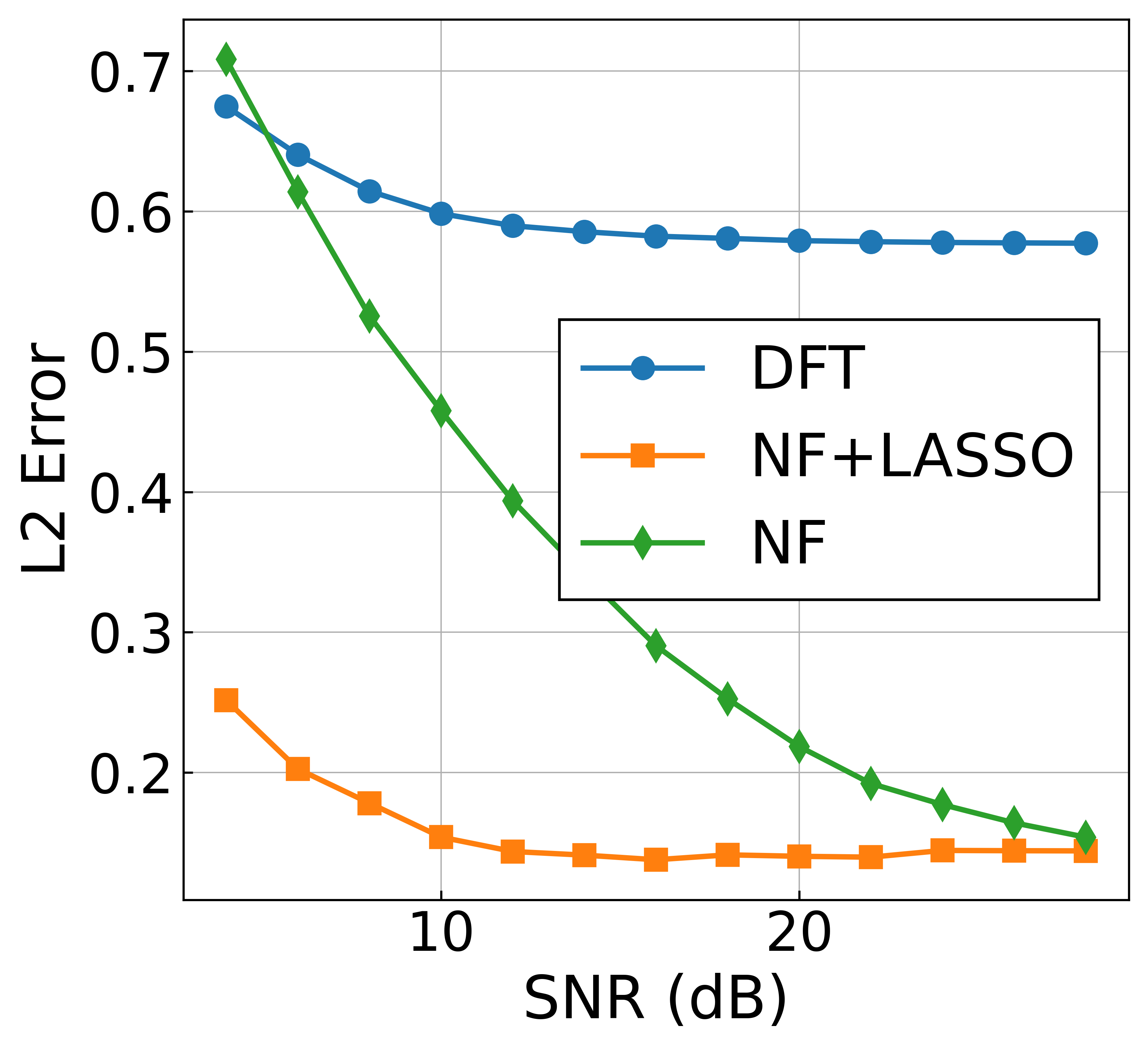}%
    \label{fig:snr_l2}}%
  \quad
  \subfloat[\tiny(b)][\textrm{\small The achievable rate versus SNR.}]{%
    \includegraphics[width=0.45\linewidth]{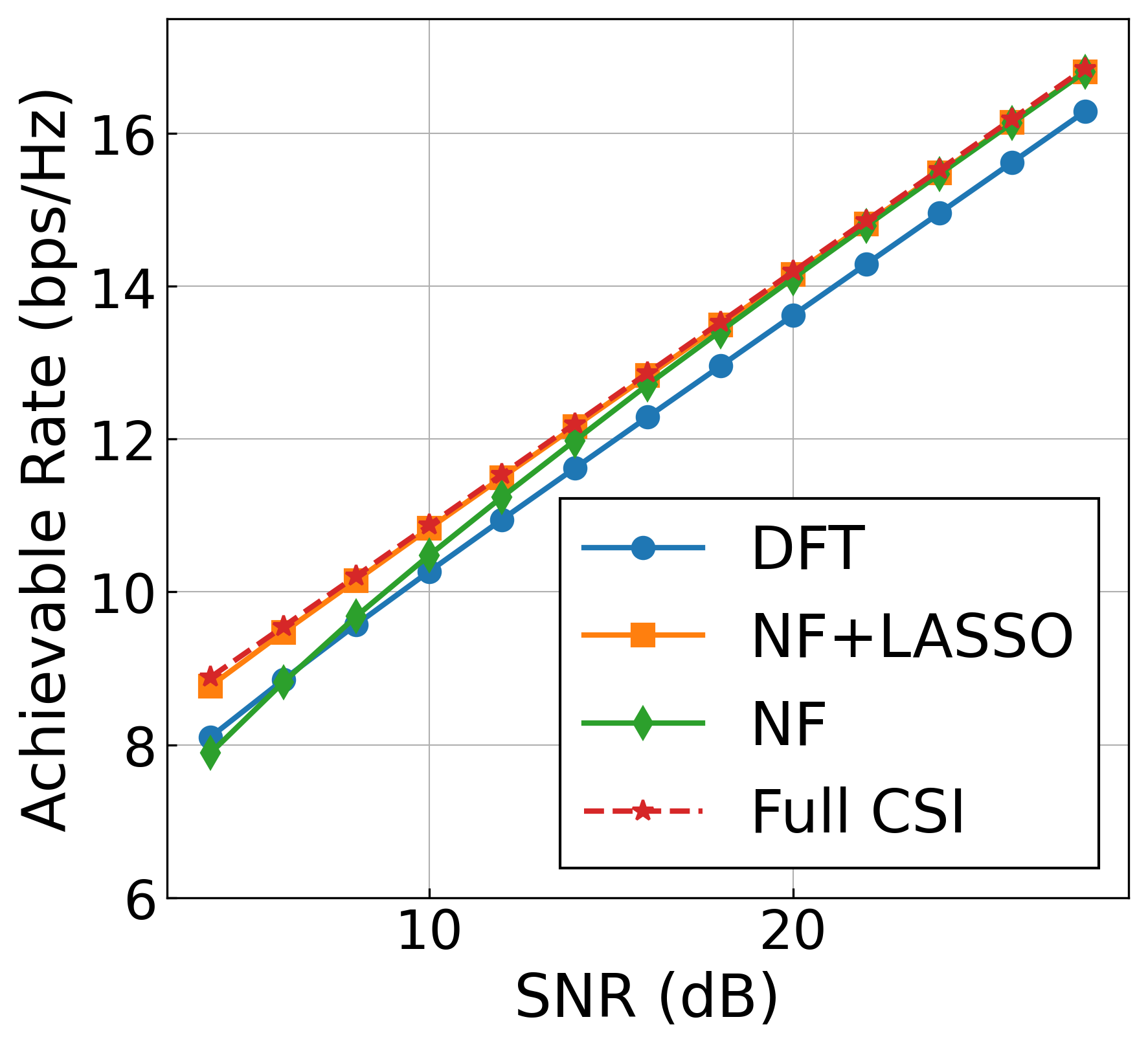}%
    \label{fig:snr_rate}}%
  \captionsetup{justification=raggedright, singlelinecheck=false} 
  \caption{Influence of SNR with 15 feedback overhead number.}
  \label{fig:snr}
\end{figure}

\begin{figure}[!t]
  \centering
  \subfloat[\tiny(a)][\textrm{\small The L2 error versus SNR.}]{%
    \includegraphics[width=0.45\linewidth]{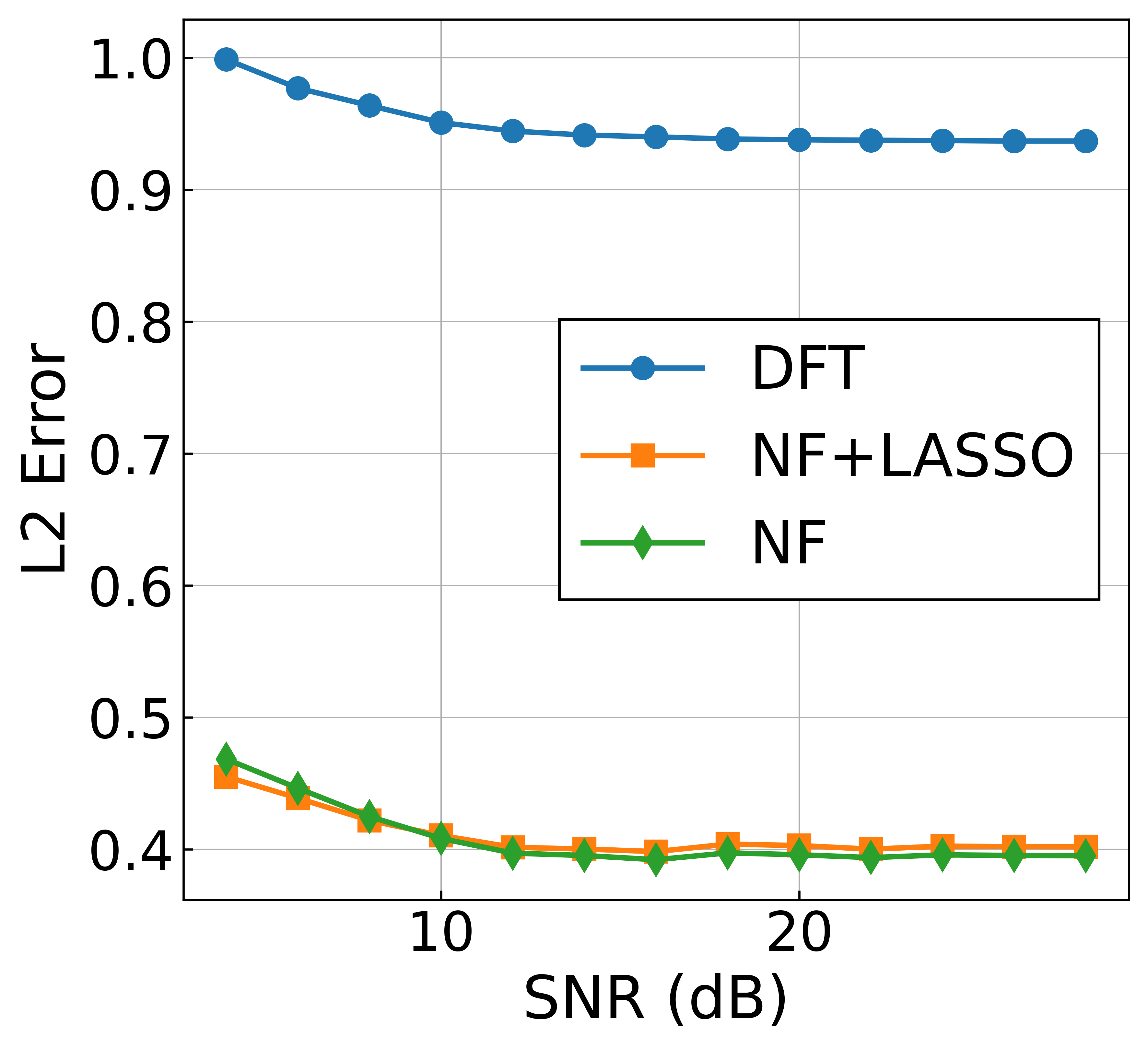}%
    \label{fig:snr_l2_5}}%
  \quad
  \subfloat[\tiny(b)][\textrm{\small The achievable rate versus SNR.}]{%
    \includegraphics[width=0.45\linewidth]{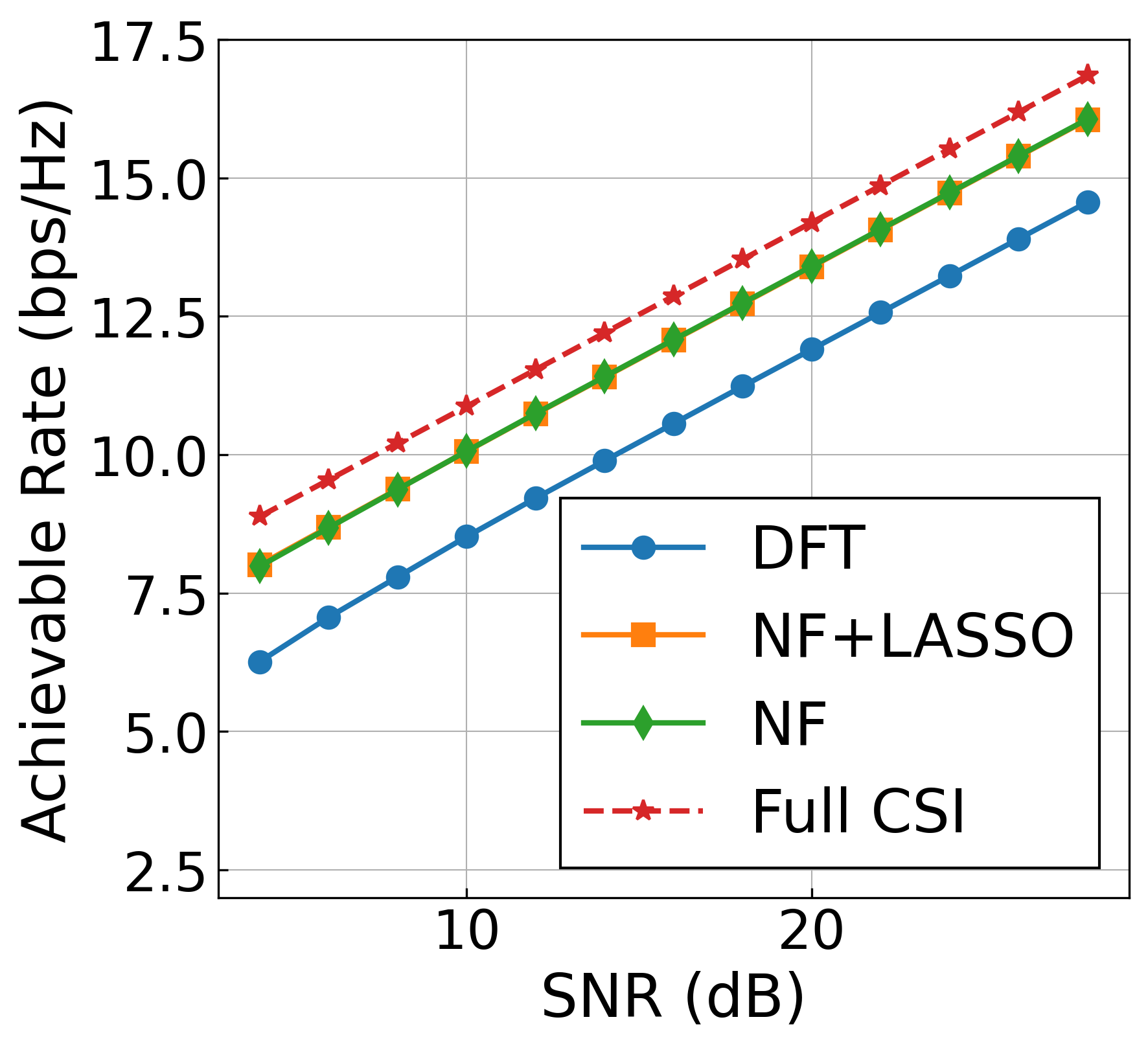}%
    \label{fig:snr_rate_5}}%
  \captionsetup{justification=raggedright, singlelinecheck=false} 
  \caption{Influence of SNR with 5 feedback overhead number.}
  \label{fig:snr_5}
\end{figure}

Next, we examine the influence of multipath effects. To isolate the impact of multipath propagation from noise, we set the SNR to 40 dB and vary the number of paths $L$ from 1 (corresponding to a line-of-sight scenario) up to 9. The resulting performance metrics are detailed in \Cref{fig:path_overhead}, which presents the minimum required feedback overhead to achieve a nearly perfect rate performance (i.e., 99\% of the achievable rate attained with true CSI) as a function of the number of paths $L$. While the required overhead increases with the number of paths for all schemes, those employing the near-field codebook consistently outperform the DFT scheme, reducing the necessary feedback overhead by up to 43.75\%. 
\vspace{-1mm}
Then, we fix the feedback overhead number at 15 and examine the influence of SNR by varying it from 4 dB to 30 dB; the corresponding results are presented in \Cref{fig:snr}. As illustrated, the NF + LASSO scheme delivers the best performance across all SNR values. Meanwhile, the NF scheme outperforms the DFT scheme when the SNR exceeds 10 dB, and its performance converges to that of the NF + LASSO scheme at high SNR levels (above 25 dB in our settings). Additionally, the performance gap between all three schemes and the upper bound represented by the red dashed curve (i.e., the scenario with perfect CSI) diminishes as SNR increases. As we reduce the feedback overhead to 5, as shown in \Cref{fig:snr_5}, both the NF and NF + LASSO schemes consistently deliver superior performance across all SNR values, while the DFT scheme exhibits a degradation of approximately 1.5 bps/Hz in achievable rate compared with the other two schemes.
\begin{figure}[!t]
  \centering
  \subfloat[\tiny(a)][\textrm{\small The L2 error versus overhead.}]{%
    \includegraphics[width=0.45\linewidth]{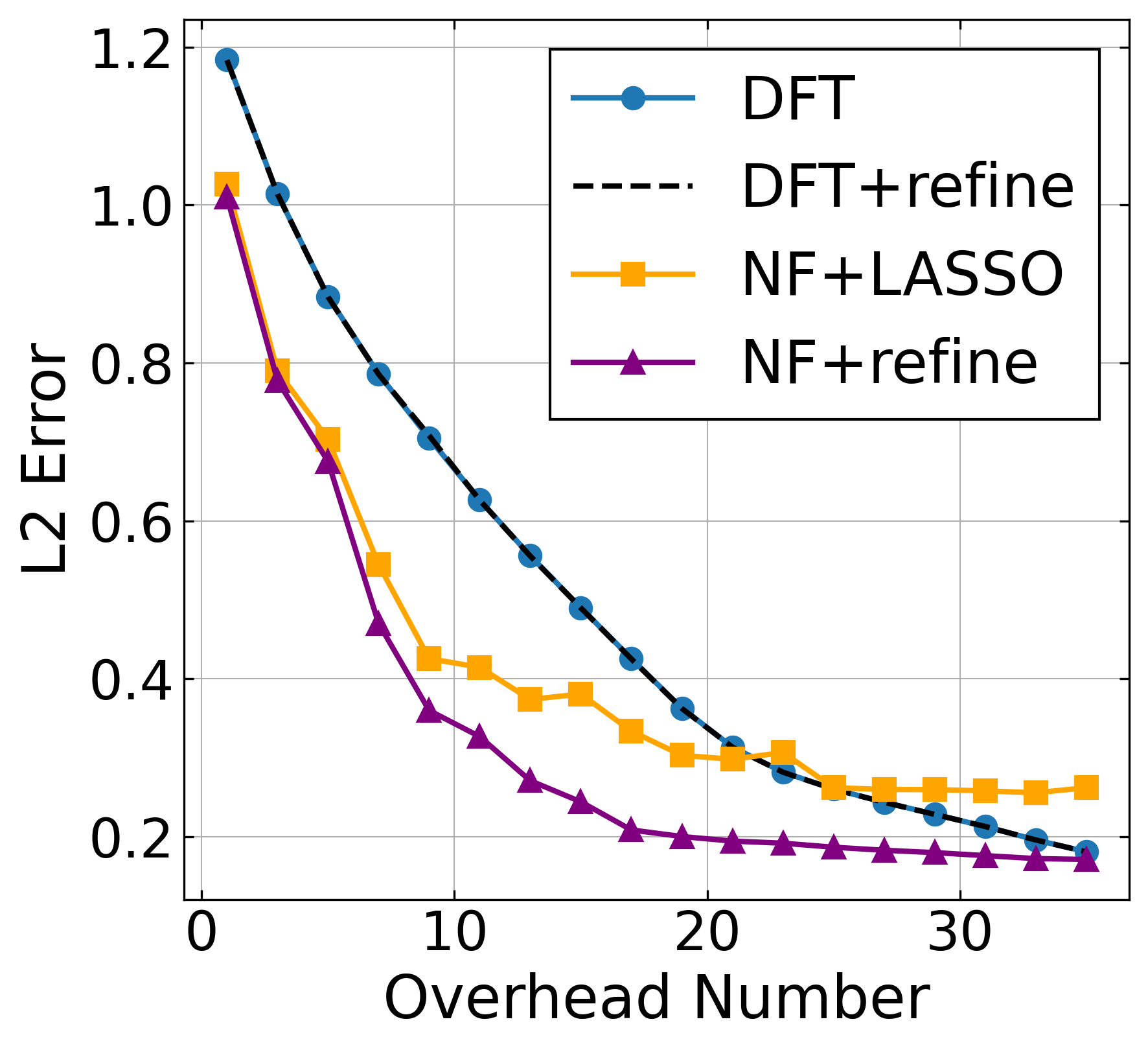}%
    \label{fig:l2_onoff}}%
  \quad
  \subfloat[\tiny(b)][\textrm{\small The achievable rate versus overhead.}]{%
    \includegraphics[width=0.45\linewidth]{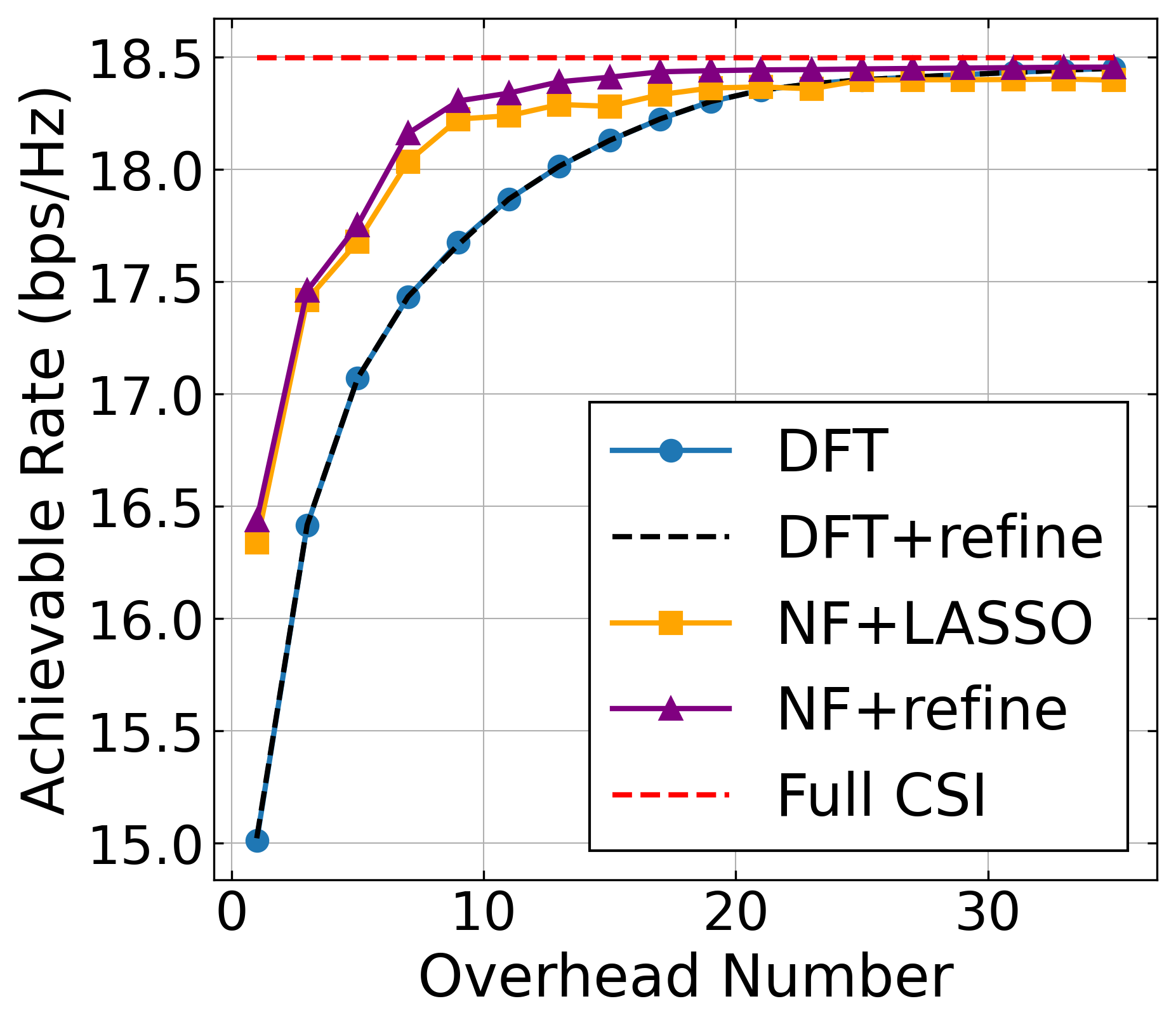}%
    \label{fig:rate_onoff}}%
  \captionsetup{justification=justified, singlelinecheck=false} 
  \caption{Performance versus feedback overhead under grid mismatches (Near-field codebook size: 1890; DFT codebook size: 512.}
  \label{fig:on_off}
\end{figure}
\section{Off-grid Refined Scheme}\label{sec:RefinedLASSO}
While these baseline schemes represent the channel as a linear combination of discrete beam-sweeping codewords, on-grid sampling inherently limits accuracy when user/scatterer angles or distances deviate from predefined grid points, particularly under constrained codebook sizes, where coarse sampling grids exacerbate discretization errors. To overcome this limitation, we propose an off-grid refinement scheme that optimizes angle-distance parameters as continuous variables. Denoting the entries in the selected index set $\mathcal{K}=\{k_1,k_2,\cdots,k_{K}\}$, the resulting optimization problem is formulated as follows:
\begin{equation}
\min_{\boldsymbol{\alpha}_{\mathcal{K}},\mathbf{\theta},\mathbf{r}}  \frac{1}{2}\left\|\mathbf{y}_{\mathcal{K}}^H - \mathbf{V}_{\mathcal{K}}^H\mathbf{V}_{\mathcal{K}}(\boldsymbol{\theta},\mathbf{r})\boldsymbol{\alpha}_{\mathcal{K}}\right\|_2^2 + \lambda \|\boldsymbol{\alpha}_{\mathcal{K}}\|_1,
    \label{eq:lasso_refined}
\end{equation}
where $\boldsymbol{\theta}=[\theta_{k_1},\theta_{k_2},\cdots,\theta_{k_{K}}],\,\mathbf{r}=[r_{k_1},r_{k_2},\cdots,r_{k_{K}}]$ are vectors of angles and distances parameterizing the codebook $\mathbf{V}_{\mathcal{K}}(\boldsymbol{\theta},\mathbf{r})=[\mathbf{b}'(\theta_{k_1},r_{k_1}),\cdots,\mathbf{b}'(\theta_{k_{K}},r_{k_{K}})]$. By optimizing these parameters alongside the sparse weights $ \boldsymbol{\alpha}_{\mathcal{K}}$, the scheme mitigates discretization errors inherent to on-grid codebooks. The problem is solved iteratively via gradient descent. 

Finally, the estimated near-field channel is reconstructed as
\begin{equation}
    \hat{\mathbf{h}} = \mathbf{V}_{\mathcal{K}}(\hat{\boldsymbol{\theta}}, \hat{\mathbf{r}})\,\hat{\boldsymbol{\alpha}}^\text{refine}_{\mathcal{K}},
    \label{eq:channel_est_refined}
\end{equation}
where \(\hat{\boldsymbol{\theta}}\), \(\hat{\mathbf{r}}\), and \(\hat{\boldsymbol{\alpha}}^\text{refine}_{\mathcal{K}}\) denote the optimized parameters in \Cref{eq:lasso_refined}. We denote the off-grid refined scheme using the near-field codebook as NF + refine.

For DFT codebooks, we substitute 
$\mathbf{V}_{\mathcal{K}}({\boldsymbol{\theta}}, {\mathbf{r}})$ with $\mathbf{V}_{\mathcal{K}}({\boldsymbol{\theta}})$ in \Cref{eq:lasso_refined} and remove the LASSO regularization term to formulate the optimization problem
\begin{equation}
\min_{\boldsymbol{\alpha}_{\mathcal{K}},\mathbf{\theta}}  \frac{1}{2}\left\|\mathbf{y}_{\mathcal{K}}^H - \mathbf{V}_{\mathcal{K}}^H\mathbf{V}_{\mathcal{K}}(\boldsymbol{\theta})\boldsymbol{\alpha}_{\mathcal{K}}\right\|_2^2,
    \label{eq:dft_refined}
\end{equation}
 and we only consider \(\boldsymbol{\theta}\) and \(\boldsymbol{\alpha}_{\mathcal{K}}\) as optimization variables. This refined scheme is referred to as DFT + refine.

\begin{figure}[!t]
  \centering
  \subfloat[\tiny(a)][\textrm{\small The L2 error versus overhead.}]{%
    \includegraphics[width=0.45\linewidth]{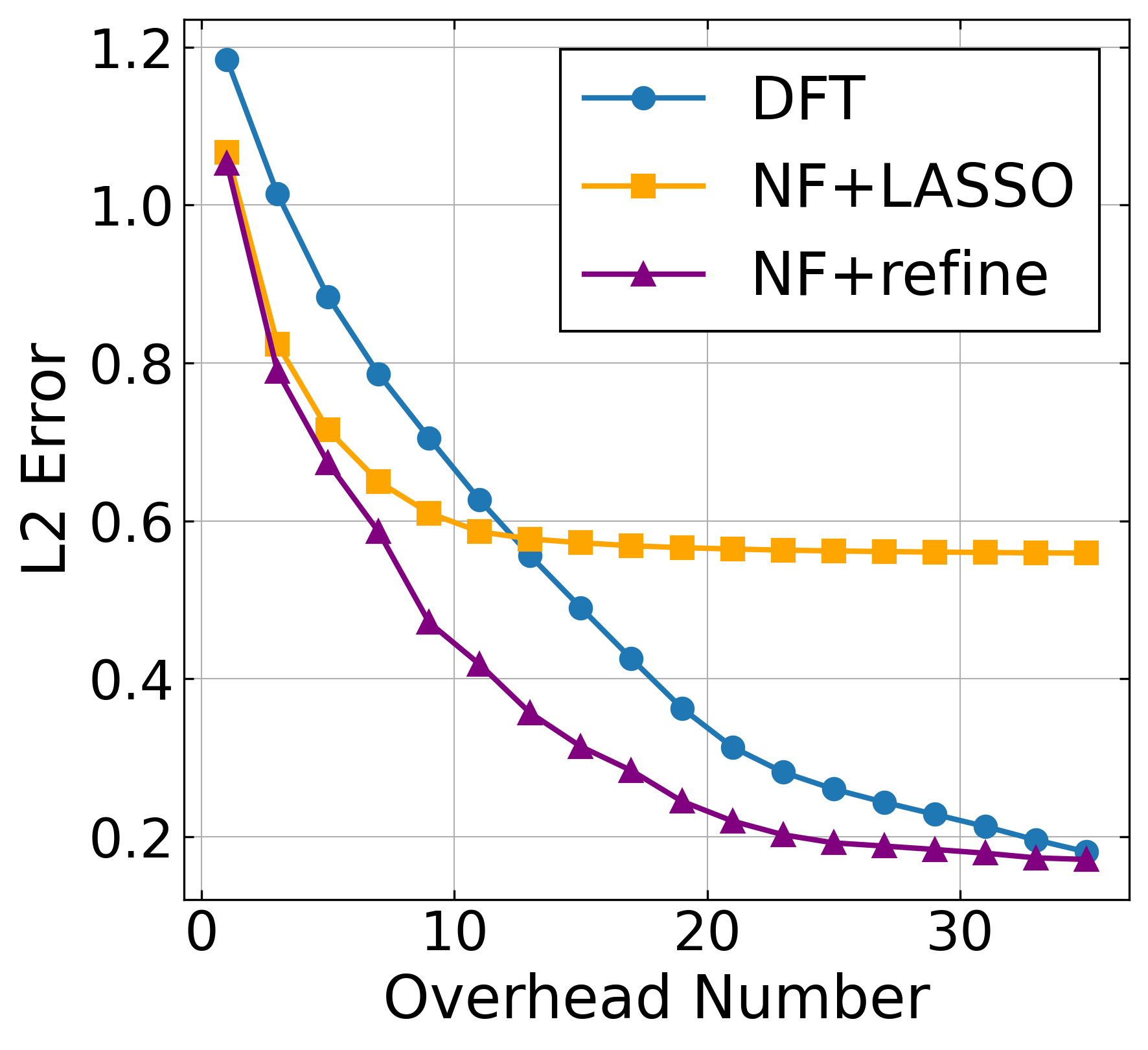}%
    \label{fig:refine_error}}%
  \quad
  \subfloat[\tiny(b)][\textrm{\small The achievable rate versus overhead.}]{%
    \includegraphics[width=0.45\linewidth]{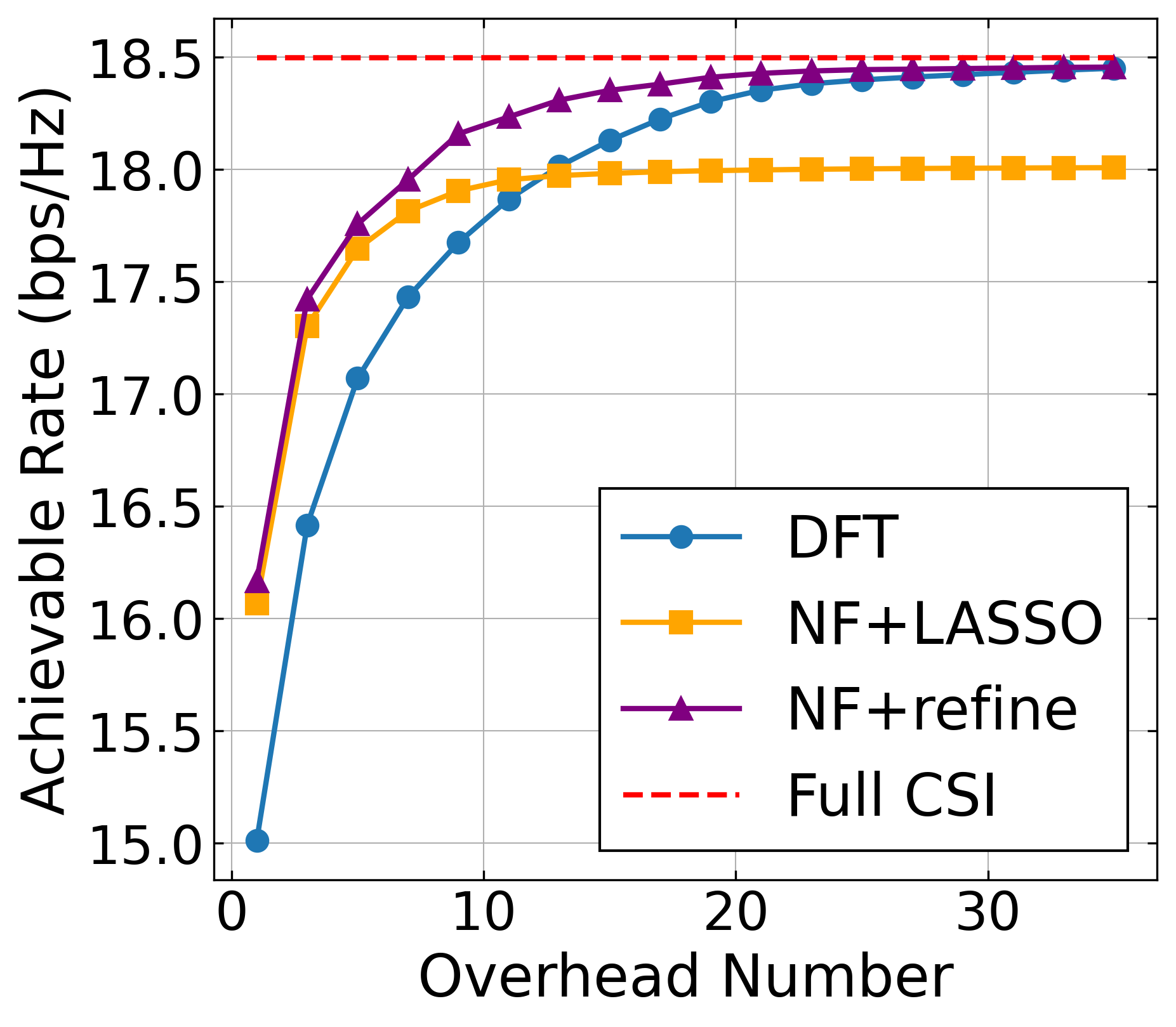}%
    \label{fig:refine_rate}}%
  \captionsetup{justification=justified, singlelinecheck=false} 
  \caption{Performance versus feedback overhead under grid mismatches for different schemes (Near-field codebook size: 520; DFT codebook size: 512).}
  \label{fig:refine}
\end{figure}

We then evaluate the performance of different schemes under grid mismatch conditions—where user and scatterer positions deviate from predefined grid points—using varying codebook sizes in \Cref{fig:on_off,fig:refine}. As detailed in \Cref{sec:simulation}, the DFT codebook employs 512 codewords, while the NF + LASSO scheme utilizes 1,890 codewords to accommodate joint angular and distance sampling. \Cref{fig:on_off} compares the L2 reconstruction error and achievable rate versus feedback overhead for both on-grid and off-grid refined schemes, with SNR fixed at 30 dB.  The DFT and DFT + refine schemes exhibit nearly identical performance due to the fine sampling in the angular domain. In contrast, the achievable rate of the NF + LASSO scheme shows a slight degradation (less than 0.01 bps/Hz) compared to the NF + refine scheme. This discrepancy is attributed to the additional distance-domain sampling in the near-field codebook, which makes it more sensitive to grid mismatches than the DFT codebook. Results indicate that grid mismatch effects are minimal for both codebooks due to their fine-grained sampling grids. However, the near-field codebook inherently requires a larger size (1,890 vs. 512 codewords) to maintain equivalent angular resolution while incorporating distance-domain sampling.

To align the NF + LASSO codebook size with the DFT baseline, we reduce its angular and distance sampling resolution, resulting in a compact near-field codebook of 520 codewords ($\beta=2.384$, sampling 332 angles and at most 2 distances for each angles). \cref{fig:refine} demonstrates that this coarser grid amplifies sensitivity to grid mismatches, causing a 2.3\% degradation in achievable rate (from 18.45 bps/Hz to 18.01 bps/Hz). However, applying the off-grid refinement strategy (see purple line in \Cref{fig:refine}) mitigates these effects: channel reconstruction accuracy improves by 69.4\%, and the achievable rate recovers to 18.46 bps/Hz, surpassing both the DFT codebook and the NF + LASSO scheme. 
\section{Conclusion}\label{sec:clun}
This paper tackles the challenges of near-field beam training in multipath environments by introducing a multi-beam linear combination framework. By adaptively combining multiple beams based on user-side feedback of the received signal's amplitude and phase, the proposed method effectively captures significant propagation paths. To address the over-completeness of the near-field codebook, a sparsity-aware formulation based on LASSO is employed, which simultaneously identifies dominant paths and suppresses noise effects in low-SNR regimes. Additionally, an iterative optimization strategy is introduced to correct grid mismatches and enhance channel estimation accuracy, particularly when the codebook size is limited.  Numerical results indicate that the near-field codebook can achieve higher rates under lower feedback overheads when compared to the DFT codebook.

\printbibliography


\end{document}